%% file: paper.tex
\documentclass[sigconf,natbib=true]{acmart}

\AtBeginDocument{%
 \providecommand\BibTeX{{%
 \normalfont B\kern-0.5em{\scshape i\kern-0.25em b}\kern-0.8em\TeX}}}

\usepackage{makecell}
\usepackage{rotating}
\usepackage{multirow}
\usepackage{lscape}
\usepackage{color, colortbl}
\usepackage{MnSymbol}

\definecolor{Gray}{gray}{0.9}
\definecolor{White}{rgb}{255,255,255}

\newboolean{showcomments}
\setboolean{showcomments}{true} 
\ifthenelse{\boolean{showcomments}}
 {
 
 }
 {\newcommand{\nbb}[2]{}
 
 }

\setcopyright{acmcopyright}
\copyrightyear{2022}
\acmYear{2022}
\acmDOI{XXXXXXX.XXXXXXX}

\acmConference[ESEM'22]{ESEM}{2022}{Helsinki, Finland}
\acmPrice{15.00}
\acmISBN{978-1-4503-XXXX-X/18/06}

\begin{document}


\title{Supporting the Task-driven Skill Identification in Open Source Project Issue Tracking Systems} 




\author{Fabio Santos}
\affiliation{%
 \institution{Northern Arizona University}
 \country{United States} \\
 \city{fabio\_santos@nau.edu}
}

\renewcommand{\shortauthors}{Fabio Santos}

\begin{abstract}


\textbf{[Background]} 
Selecting an appropriate task is challenging for contributors to Open Source Software (OSS), mainly for those who are contributing for the first time. Therefore, researchers and OSS projects have proposed various strategies to aid newcomers, including labeling tasks. 
\textbf{[Aims]} 
In this research, we investigate the automatic labeling of open issues strategy to help the contributors to pick a task to contribute. We label the issues with API-domains---categories of APIs parsed from the source code used to solve the issues. We plan to add social network analysis metrics gathered from the issues conversations as new predictors. By identifying the skills, we claim the contributor candidates should pick a task more suitable to their skill.
\textbf{[Method]} 
We are employing mixed methods. We qualitatively analyzed interview transcripts and the survey's open-ended questions to comprehend the strategies communities use to assist  in onboarding contributors and contributors used to pick up an issue. We applied quantitative studies to analyze the relevance of the API-domain labels in a user experiment and compare the strategies' relative importance for diverse contributor roles. We also mined project and issue data from OSS repositories to build the ground truth and predictors able to infer the API-domain labels with comparable precision, recall, and F-measure with the state-of-art. We also plan to use a skill ontology to assist the matching process between contributors and tasks. By quantitatively analyzing the confidence level of the matching instances in ontologies describing contributors' skills and tasks, we might recommend issues for contribution. In addition, we will measure the effectiveness of the API-domain labels by evaluating the issues solving time and the rate among the labeled and unlabelled ones. 
\textbf{[Results]} 
So far, the results showed that organizing the issues--which includes assigning labels is seen as an essential strategy for diverse roles in OSS communities. The API-domain labels are relevant, mainly for experienced practitioners. The predicted labels have an average precision of 75.5\%. 
\textbf{[Conclusions]} 
Labeling the issues with the API-domain labels indicates the skills involved in an issue. The labels represent possible libraries (aggregated into domains) used in the source code related to an issue. By investigating this research topic, we expect to assist the new contributors in finding a task, helping OSS communities to attract and retain more contributors.
\end{abstract}


\maketitle

\section{Context and Problem}
\label{sec:context_problem}


The first steps toward contributing to OSS are challenging for some developers \cite{steinmacher2015systematic}. Developers are often required to pick up a task from a list of open issues, which may have varying complexity levels and require different skills to be completed. The skills required for a task are hard to presume based on the available task data~\citep{Bettenburg:2007:QBR:1328279.1328284,vaz2019empirical,zimmermann2010makes}. The difficulty in identifying skills in the context of software development might be caused by the diversity of knowledge concepts. They can range from certain program concepts to a more fine-grained level (even a specific class or package name) and encompass hard or soft skills~\citep{vadlamani2020studying,fitts1967human}.

Adding labels to the issues (a.k.a tasks, bug reports) helps  contributors when they are choosing their tasks~\cite{steinmacher2018let}. However, community managers find labeling issues challenging and time-consuming \cite{9057411} because projects require skills in different languages, frameworks, database management systems, and Application Programming Interfaces (APIs). Many recent studies point to manual or automatic methods to classify issues, but the classification is restricted to the nature of the issue---bug/non-bug, questions, documentation, etc.~\citep{antoniol2008bug,fan2017road,kallis2019ticket,otoom2019automated,pandey2017automated,pingclasai2013classifying,xia2013tag,zhou2016combining,zhu2019bug}. 

Issue Tracking Systems (e.g., GitHub, Jira) create an environment that enables collaboration among developers. Previous studies leverage ``social metrics'' calculated from data mined from the developers interactions to compose successful predictive models in diverse domains: co-changes \cite{wiese2017using}, defects \cite{zimmermann2008predicting} and failures \cite{meneely2008predicting}. \citet{wiese2014social} studied which metrics were used as predictors. In this research, we aim to explore whether such metrics can improve our predictive models. Uncovering good predictors can also help us hypothesize about conceptual relationships between socials aspects of the software development and skill identification.

In this work, we extend the existing research by proposing an automated approach to identify skills required to work on an issue.  We apply machine learning and a skill ontology
to assist in matching the issues' and new contributors' skills. The skills predicted will be presented as labels based on categories of APIs, called API-domain labels (``UI'', ``Cloud'', ``Error Handling'' etc), defined by a group of experts. APIs usually encapsulate modules with specific purposes (e.g., cryptography, database access, logging, etc.), abstracting the underlying implementation. If the contributors know which categories of APIs will be required to work on each issue, they could choose tasks that better match their skills or involve skills they want to learn. 

We expect that our work will have the following contributions: 

\begin{itemize}

\item definition of categories or API-domain labels to be used in software projects using a skill ontology.

\item an approach to predict labels using the information related to the issue.

\item an investigation of the usefulness of the API-domain labels.

\item an evaluation of the labeling skills as a strategy with diverse stakeholders.




\end{itemize} 


\section{Research Goals and Questions}


To address the aforementioned problem, we propose the following research questions:


\textbf{RQ1.} To what extent can we predict the domain of APIs used in the code that fixes a software issue?

RQ1 aims to propose and evaluate an automatic labeling approach quantitatively.




\textbf{RQ2.} How does the labeling strategy impact the issue choice?

RQ2 aims to investigate the label's strategy and the API-domain labels from the point of view of communities and contributors.









\textbf{RQ3.} To what extent can contributors match their skills with tasks?

RQ3 aims to evaluate the possibility of automatically matching the issues and tasks' skills. 



By understanding how new and experienced contributors select the issues on ITS 
 and the strategies communities use to assist contributors, it is possible to create an approach to identify, label the skills and match with candidate contributors. 
Therefore, helping them choose the most appropriate task and receive feedback about the labels generated.

\section{Research Plan}

\begin{figure*}[htb]
\centering
\includegraphics[width=.8\textwidth]{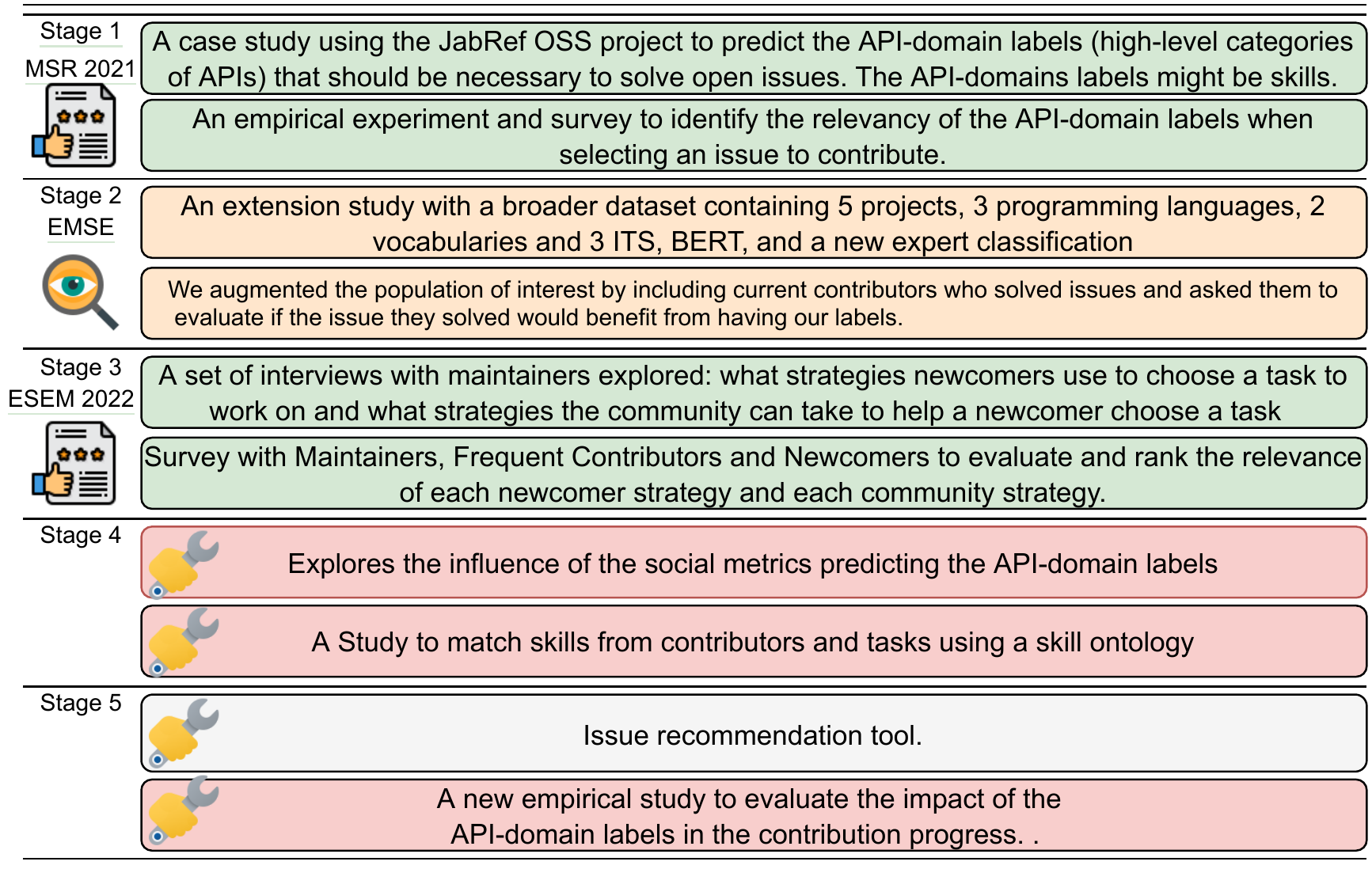}
\caption{Research Method Overview. Green = published, Yellow = under review, Red = in progress, Gray = tool.}
\label{fig:reserachMethodOverview}
\end{figure*}

The study started with a case study (JabRef project), which identified skills based on the API-Domains mined from GitHub. The API labels were predicted with precision, recall, and F-measure of 0.755, 0.747, and 0.751, respectively using the Random Forest algorithm~\citep{santos2021can} (Figure \ref{fig:reserachMethodOverview} - Stage 1).

We also ran a user experiment with 74 students and practitioners to assess the labels' relevancy. The participants picked a task in a mocked page, in which we inserted API-domain labels. They also reported which regions of the open issue page they perceived as relevant. Finally, they explained why the issue regions were relevant and what type of labels they would like to see in an ITS~\citep{santos2021can} (Figure \ref{fig:reserachMethodOverview} - Stage 1). The participants found the labels information important to decide which issue to contribute and the API-domains labels relevant even more than the components labels which are present in the project. 

Then, we interviewed 17 maintainers of the projects to learn \textit{``how do newcomers choose an issue, and how can the community help?''}. 
(Figure \ref{fig:reserachMethodOverview} - Stage 3). We also investigated mismatches of perceptions from diverse OSS community stakeholders regarding how to help new contributors find a task to start with. The strategies' investigation counted with a survey with 64 participants who evaluated the newcomers' and maintainers' strategies proposed by the 17 interviewees. The study concluded new contributors and maintainers disagree about the relative importance of the onboarding newcomers community's strategy and new contributors and frequent contributors disagree about the importance of the setup the environment contributors' strategy. The study will appear in ESEM 2022 \cite{santos2022how}.


We are extending the tooling to identify skills employed in the case to use a model trained with BERT, three different ITS: GitHub, Gerrit, and Jira, three programming languages: Java, C++, and C\#, five projects: JabRef, audacity, PowerToys, RTTS, and Cronos. Those two last are projects from the industry. Extending to industry projects aims to verify whether stakeholders can apply the research to industry software. The improvements embraced a new semi-automatic API classification model carried out manually in the first experiment (Figure \ref{fig:reserachMethodOverview} - Stage 2). We are also asking project developers to analyze the labels predicted from the new projects to give us feedback about the labels.

In the following steps, we will use the results from social network analysis to predict the API-domain labels. The API-domain labels will use a skill ontology to match users' and tasks' skills in an experiment (Figure \ref{fig:reserachMethodOverview} - Stage 4). Finally, the API-domain labels will be evaluated by applying them to random issues. After a period, we will assess the percentage of solved issues and the time of solution. At the same time, a tool will suggest issues to contribute based on the contributor's skills informed in a web form (Figure \ref{fig:reserachMethodOverview} - Stage 5). 


\section{Publications}
\label{sec:publications}

The research conducted as part of this dissertation resulted in the publications:

\underline{Santos, F.}, Trinkenreich, B., Nicolati Pimentel, J.F., Wiese, I., Steinmacher, I., Sarma, A. and Gerosa, M.A., 2022, June. How to choose a task? Mismatches in perspectives of newcomers and existing contributors. In: \textit{ 2022 16th ACM/IEEE International Symposium on Empirical Software Engineering and Measurement (ESEM) }. ACM/IEEE.

\underline{Santos, F.}, Trinkenreich, B., Wiese, I., Steinmacher, I., Sarma, A. and Gerosa, M.A., 2021, May. Can I Solve It? Identifying APIs Required to Complete OSS Tasks. In: \textit{ 2021 IEEE/ACM 18th International Conference on Mining Software Repositories (MSR) (pp. 346-257)}. IEEE.

I also participated in the CHASE 2021 conference as WebChair.

\section{Expected Contributions}

To the best of our knowledge, no previous work provides ways to automatically label the skills in OSS issues, match the contributors and tasks skills using a skill ontology, and evaluate the labels strategy and the predicted labels with users. 

\section{Main Preliminary Results } 
\label{sec:preliminary_results_stage1}

\subsection{Stage 1 - JabRef Case Study} 
\label{sec:results-jabref-case}

We report the results grouped by research question \footnote{More details about the results, including statistical tests, are available in the published papers}.

\subsubsection{RQ1. To what extent can we predict the domain of APIs used in the code that fixes a software issue?}

To answer this research question, we predicted the API-domain labels using diverse configurations of corpus (only \textsc{title} (T), only \textsc{body} (B), \textsc{title} and \textsc{body} (T, B), and \textsc{title}, \textsc{body}, and \textsc{comments} (T,B,Comments)), grams (unigrams to 4-grams) using TF-IDF and a multi-label classification to predict n API-domain labels. The ground truth is composed by APIs declared in source code files attached to commits and linked to issues. 

\input{Tables/resultsFRFM}

As Table~\ref{tab:results} shows, when we tested different inputs and compared to \textsc{Title} only, all alternative settings provided better results. We could observe improvements in terms of precision, recall, and F-measure. When using \textsc{title}, \textsc{body}, and \textsc{comments}, we reached Precision of 75.5\%, Recall of 74.7\%, and F-Measure of 75.1\%. 

\begin{figure}[!hbt]
\centering
\includegraphics[width=.5\textwidth, trim= 10px 25px 20px 20px]{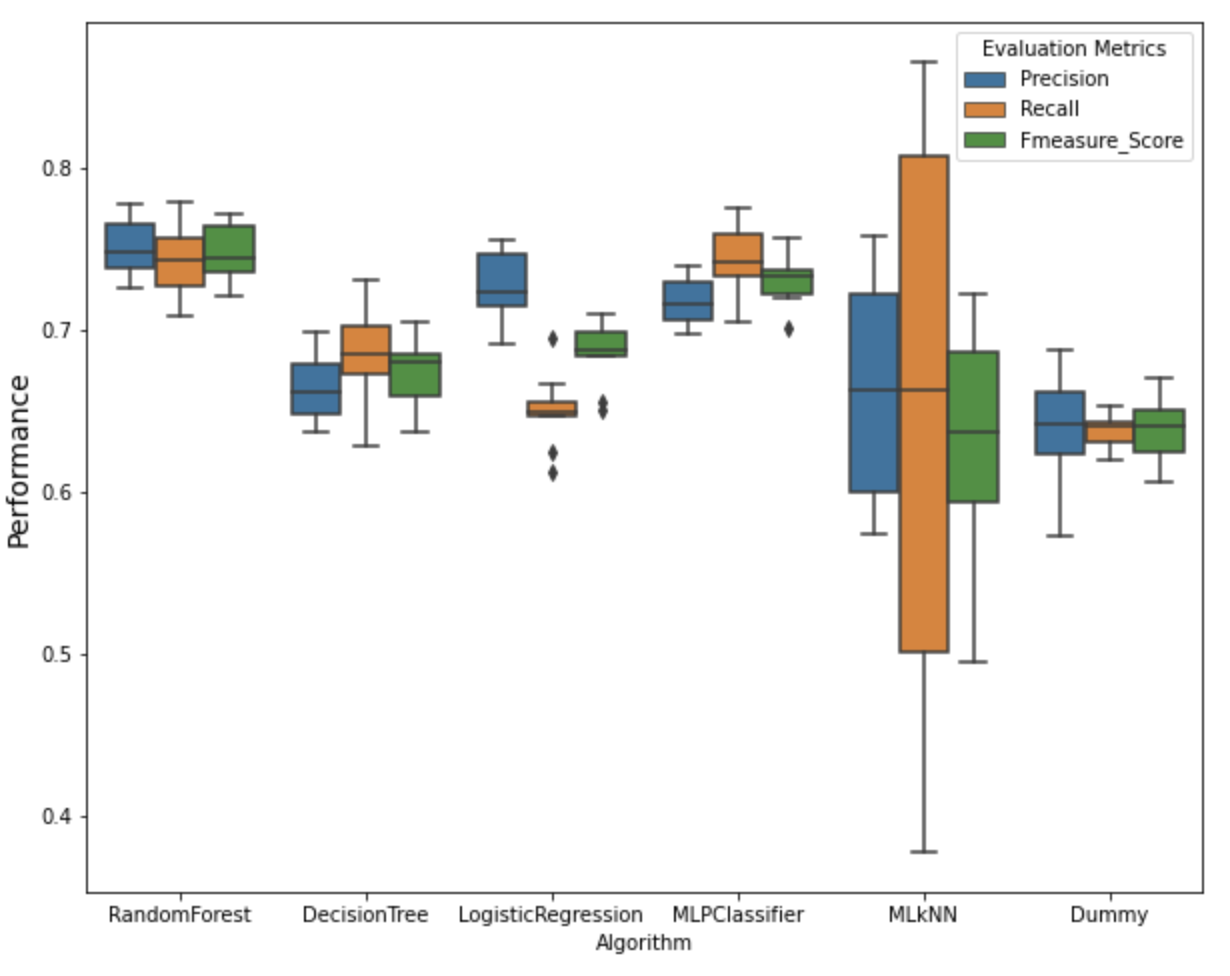}
\caption{Comparison between the baseline model and other machine learning algorithms}
\label{fig:baselineH6}
\end{figure}

Finally, to investigate the influence of the machine learning (ML) classifier, we compared several options using the title with unigrams as a corpus. The options included: Random Forest (RF), Neural Network Multilayer Perceptron (MLPC), Decision Tree (DT), LR, MlKNN, and a Dummy Classifier with strategy ``most\_frequent''. Dummy or random classifiers are often used as a baseline~\citep{saito2015precision, flach2015precision}. 
Fig. \ref{fig:baselineH6} shows the comparison among the algorithms. 

Random Forest (RF) and Neural Network Multilayer Perceptron (MLPC) were the best models when compared to Decision Tree (DT), Logistic Regression (LR), MlKNN, and Dummy algorithms. 






\subsubsection{RQ2. How relevant are the API-domain labels to new contributors?}

We conducted a user study with 74 participants to answer this research question and analyzed their responses. 

\begin{figure}[!hbt]
\centering
\includegraphics[width=.4 \textwidth] {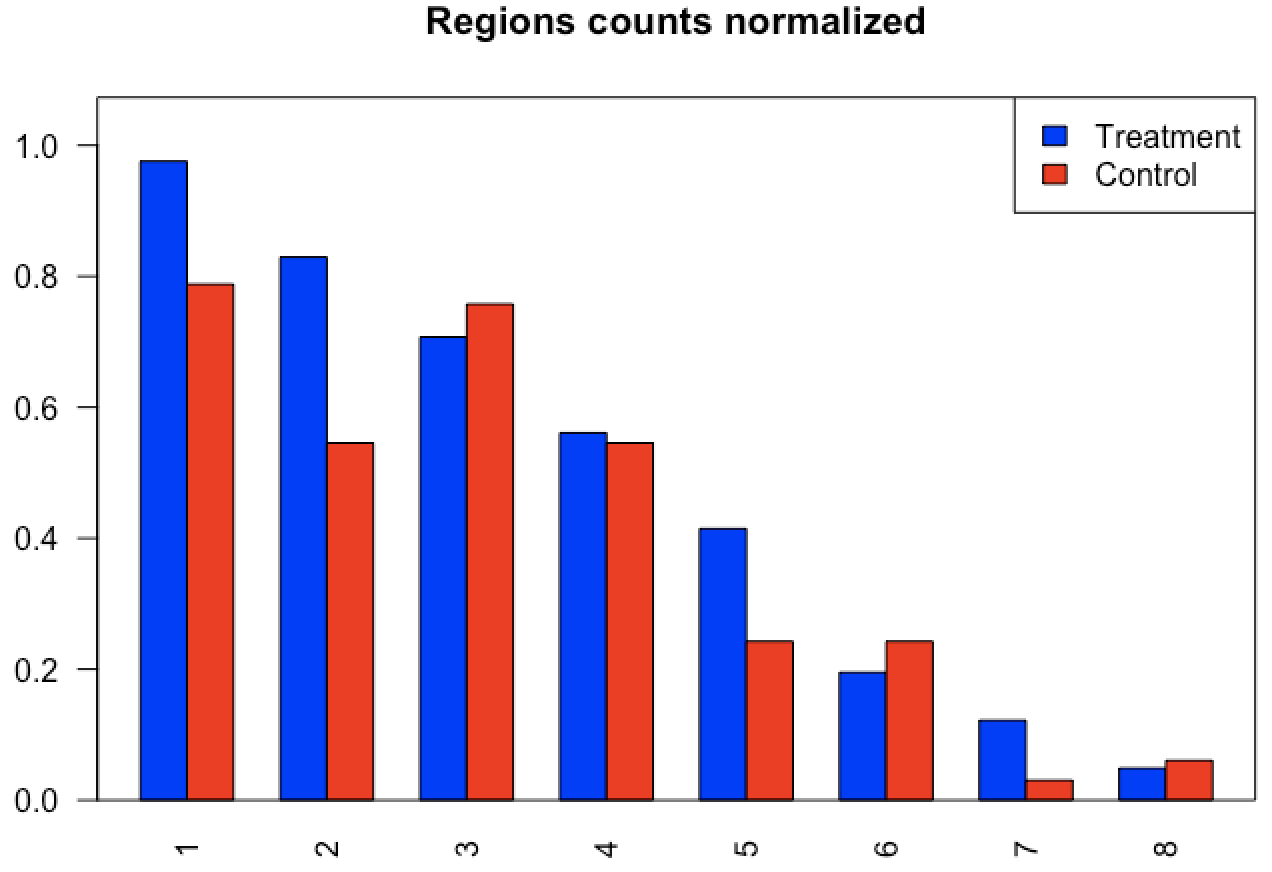}
\caption{The regions counts (normalized) of the issue's information page selected as most relevant by participants from Treatment and Control groups. 1-Title,2-Label,3-Body,4-Code,5-Comments,6-Author,7-Linked issues,8-Participants.}
\label{fig:hotmapchoicesTC}
\end{figure}

\textbf{What information is used when selecting a task?}
Understanding the type of information that participants use to decide while selecting an issue to work on can help projects better organize such details on their issue pages. Fig.~\ref{fig:hotmapchoicesTC} shows the different regions participants found useful. 


Finally, we analyzed whether the demographic subgroups had different perceptions about the API-domain labels (Table~\ref{tab:apiXcompXtype}). When comparing Industry vs. Students, we found participants from industry selected 1.9x (p-value=0.001) more API-domain labels than students when we controlled by component labels. 

The odds ratio analysis suggests that API-domain labels are more likely to be perceived relevant by practitioners and experienced developers than by students and novice coders.

\input{Tables/APixCompxType}

\input{Tables/APILabels}
\subsection{Stage 2 - Strategy Importance} 
\label{sec:strategy-importance}

\subsubsection{RQ3: What strategies help newcomers choose a task in OSS?}
\label{sec:results:rq1}

To answer this research question, we interviewed maintainers to understand their perspectives on (i) what strategies a newcomer uses to choose an open issue; and (ii) what strategies the OSS communities can use to help newcomers choose tasks. 



From the interviews, we could identify 27 strategies that maintainers expect newcomers to use to choose a task and grouped them into five categories. 
In addition to the strategies newcomers are expected to take, we found 40 strategies that the communities use to help newcomers choose their tasks. 

\subsubsection{RQ4: How do newcomers and existing contributors differ in their opinions of which strategies are important for newcomers?}
\label{sec:results:rq2}

To compare the relative importance of the strategies from the point of view of different stakeholders, we used the Schulze method~\cite{schulze2003new} to combine the rankings for (i) newcomers and (ii) community strategies. In the following subsections, we present the rankings and how they compare.

\begin{figure}[htb]
\centering
\includegraphics[width=0.9\linewidth]{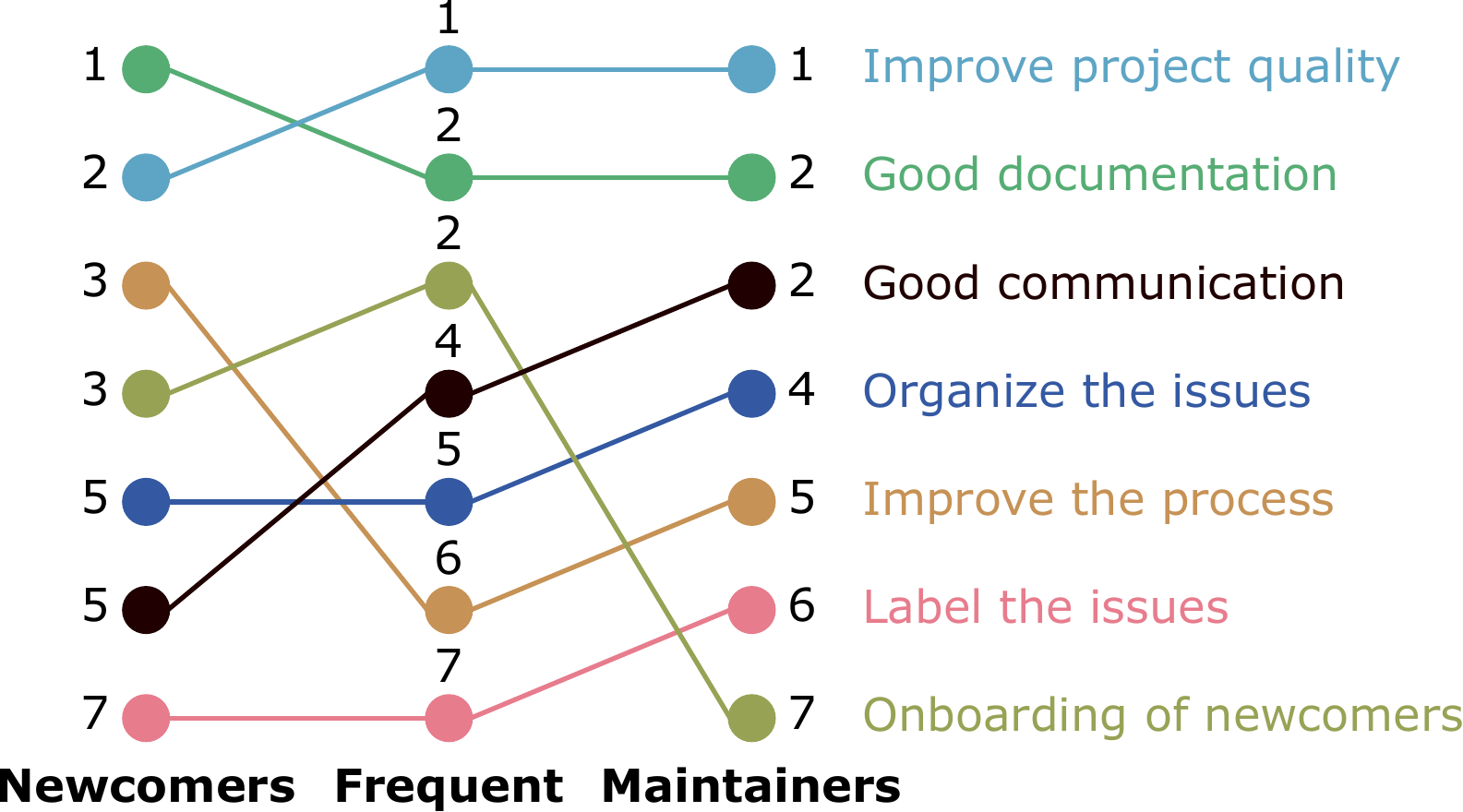}
\caption{The relative importance of community strategies}
\label{fig:strategiesflow}
\end{figure}

Fig.~\ref{fig:strategiesflow} presents the combined rankings for maintainers' strategies according to each stakeholder. Once again, the perspective of frequent contributors and maintainers are similar, with one standout difference: ``Support the onboarding of newcomers''.

Among the mismatches we found, ``Support the onboarding of newcomers,'' while it was ranked last for the maintainers, it was the third for newcomers and second for frequent contributors. 

\bibliographystyle{ACM-Reference-Format}
\bibliography{paper}

\end{document}

%% file: Tables/resultsFRFM.tex
\begin{table}[ht!]
 \begin{center}
 \caption{overall metrics from models created to evaluate the corpus. Hla - Hamming Loss }
 \label{tab:results}
 \begin{tabular}{c|r|r|r|r} 
  \hline

  \textbf{Model} & \textbf{Precision} & \textbf{Recall} & \textbf{F-measure} &
  \textbf{Hla} \\
  \hline
Title (T) 
&0.717 &	0.701 &	0.709 & 0.161\\
Body (B)
& 0.752 &	0.742 &	0.747 & 0.143\\
T, B 
& 0.751 &	0.738 &	0.744 & 0.145\\
T, B, Comments 
& 0.755	& 0.747 &	0.751 & 0.142\\
 \hline

 \end{tabular}
 \end{center}
\end{table}

%% file: Tables/APixCompxType.tex
\begin{table}[h]
 \begin{center}
 \caption{Answers from different demographic subgroups regarding the API labels (API/Component/Issue Type)}
 \label{tab:apiXcompXtype}
 \begin{tabular}{c|c|r|r} 
  \hline
  \textbf{Subgroup} & \textbf{Comparison} & \textbf{API \%} & \textbf{Comp or Type \%} \\
  \hline

Industry & API/Comp & \textbf{56.0} & 44.0 \\
Students & API/Comp & 40.0 & \textbf{60.0} \\

Exp. Coders & API/Comp & \textbf{50.9} & 49.1 \\
Novice Coders & API/Comp & 41.5 & \textbf{58.5}  \\

Industry & API/issue Type & 45.5 & \textbf{55.5} \\
Students & API/issue Type & 30.6 & \textbf{69.4} \\

Exp. Coders & API/issue Type & 43.5 & \textbf{56.5} \\
Novice Coders & API/issue Type & 30.9 & \textbf{69.1}  \\

\hline
	
 \end{tabular}
 \end{center}
\end{table}